\documentclass{article}
\usepackage{spconf,amsmath,graphicx, booktabs}
\usepackage{xcolor}
\usepackage{amssymb}
\usepackage{url}
\usepackage{hyperref}

\title{Fast, High-Quality and Parameter-Efficient Articulatory Synthesis using Differentiable DSP}
\name{Yisi Liu, Bohan Yu, Drake Lin, Peter Wu, Cheol Jun Cho, Gopala Krishna Anumanchipalli}
\address{UC Berkeley\\
\texttt{louis\_liu@berkeley.edu}}

\begin{document}
%
\maketitle
\begin{abstract}
%
Articulatory trajectories like electromagnetic articulography (EMA) provide a low-dimensional representation of the vocal tract filter and have been used as natural, grounded features for speech synthesis. Differentiable digital signal processing (DDSP) is a parameter-efficient framework for audio synthesis. Therefore, integrating low-dimensional EMA features with DDSP can significantly enhance the computational efficiency of speech synthesis. In this paper, we propose a fast, high-quality, and parameter-efficient DDSP articulatory vocoder that can synthesize speech from EMA, F0, and loudness. We incorporate several techniques to solve the harmonics / noise imbalance problem, and add a multi-resolution adversarial loss for better synthesis quality. Our model achieves a transcription word error rate (WER) of $6.67\%$ and a mean opinion score (MOS) of 3.74, with an improvement of $1.63\%$ and 0.16 compared to the state-of-the-art (SOTA) baseline. Our DDSP vocoder is 4.9x faster than the baseline on CPU during inference, and can generate speech of comparable quality with only 0.4M parameters, in contrast to the 9M parameters required by the SOTA.

\end{abstract}
\begin{keywords}
Neural vocoder, articulatory synthesis, DDSP, computational efficiency, parameter-efficient, high-quality
\end{keywords}
\section{Introduction}
Articulatory synthesis is the task of generating speech audio from articulatory features, i.e., the physical movements of human articulators, often measured as electromagnetic articulography (EMA). Since the articulatory features are physically grounded \cite{goldstein}, EMA-to-speech vocoders are more interpretable than mel-spectrogram-based vocoders \cite{Peter-ATS}. Articulatory vocoders are also highly controllable, allowing for nuanced adjustments in speech generation \cite{Peter-ATS, articulatory_encodec}. Given these unique characteristics, articulatory synthesis has many applications including helping patients with vocal cord disorders communicate better \cite{EMA2S, dysarthric}, decoding brain signals to speech waveforms \cite{nature2019}, and augmenting silent speech systems \cite{EMA2S}. 

However, to our knowledge, there has been little investigation into the parameter efficiency of articulatory synthesis models, which is important for applications on edge devices, where the memory and computation are limited. Smaller models may also have faster inference speed, which also opens up new possibilities for faster real-time applications. Since articulatory synthesis is mostly utilized in clinical domains, a high-speed low-footprint synthesis model is crucial for maximizing accessibility.  

We utilize differentiable digital signal processing (DDSP) \cite{DDSP} to achieve efficient articulatory synthesis while maintaining high-fidelity audio generation. A DDSP model consists of a neural network encoder and traditional digital signal processing (DSP) modules. The encoder transforms input features, such as F0, loudness, and spectral features, into control signals like filter coefficients and harmonic amplitudes. DSP modules then generate audio from these control signals. The differentiability of DSP modules allows for end-to-end training, hence the term ``Differentiable DSP". DDSP models are light-weight since they utilize the strong inductive bias of known signal-processing modules to explicitly model the speech generation process \cite{ddspreview}. Consequently, DDSP models only need to learn control signals rather than raw waveforms, delegating synthesis to DSP modules. 

In this paper, we introduce a novel articulatory synthesis approach using DDSP with the Harmonic-plus-Noise (H+N) model to convert articulatory features (EMA, F0, loudness) into speech. To our knowledge, this is the first application of DDSP to articulatory synthesis. Our model achieves a word error rate (WER) of 6.67\% and a mean opinion score (MOS) of 3.74, improving the state-of-the-art (SOTA) result by 1.63\% and 0.16, respectively. It is also 4.9x faster during CPU inference. Additionally, a 0.4M parameter version of our model matches the quality and intelligibility of the previous 9M-parameter SOTA. Codes and audio samples are available at \href{https://tinyurl.com/ddsp-vocoder}{tinyurl.com/ddsp-vocoder}.

\section{Related Work}
\subsection{Articulatory Synthesis}
Articulatory synthesis with traditional digital signal processing methods has long been investigated \cite{traditional_0, traditional_1, traditional_4, vocaltract_lab}. In the deep learning era, there are generally three methods for articulatory synthesis: (1) predicting the acoustic parameters first and then using traditional signal-processing-based vocoders, e.g. WORLD \cite{world}, to synthesize speech \cite{orientation, bilstm}; (2) predicting intermediate spectrograms and then utilizing GAN-based vocoders \cite{PWG, hifigan} to convert spectrograms to speech signals \cite{EMA2S, style}; (3) directly synthesizing speech from articulatory features with HiFi-CAR \cite{Peter-ATS, articulatory_encodec, hifigan, CAR-GAN}. Among them, \cite{Peter-ATS} is the SOTA model in terms of synthesis intelligibility and inference speed, and \cite{articulatory_encodec} extends it to a universal articulatory vocoder. However, there is still scope for improving parameter efficiency and synthesis quality. 

\subsection{Differentiable Digital Signal Processing}

There are two main architectures of DDSP synthesizers: (1) the source-filter model \cite{nsf, meta, nhv}, and (2) the Harmonic-plus-Noise (H+N) model \cite{dworld, fabbro2020speech, nercessian2021}. Since H+N models are strictly more expressive than source-filter models \cite{DDSP}, we investigate the H+N model in this paper. The H+N model divides speech into two components: harmonics, which represent the periodic part of speech produced by vocal cord vibrations; and noise, which models the aperiodic component of speech produced by airflow in the vocal tract. DDSP has wide-spread applications in music generation \cite{DDSP, NSFmusic}, timbre transfer\cite{timebre_1, timebre_3}, singing voice synthesis\cite{sing_1, sing_2}, and speech synthesis \cite{nsf, meta, nhv, dworld, fabbro2020speech, nercessian2021}.

\section{Methods}
\label{methods}
\begin{figure}[t]
    \centering
    \includegraphics[scale=0.42]{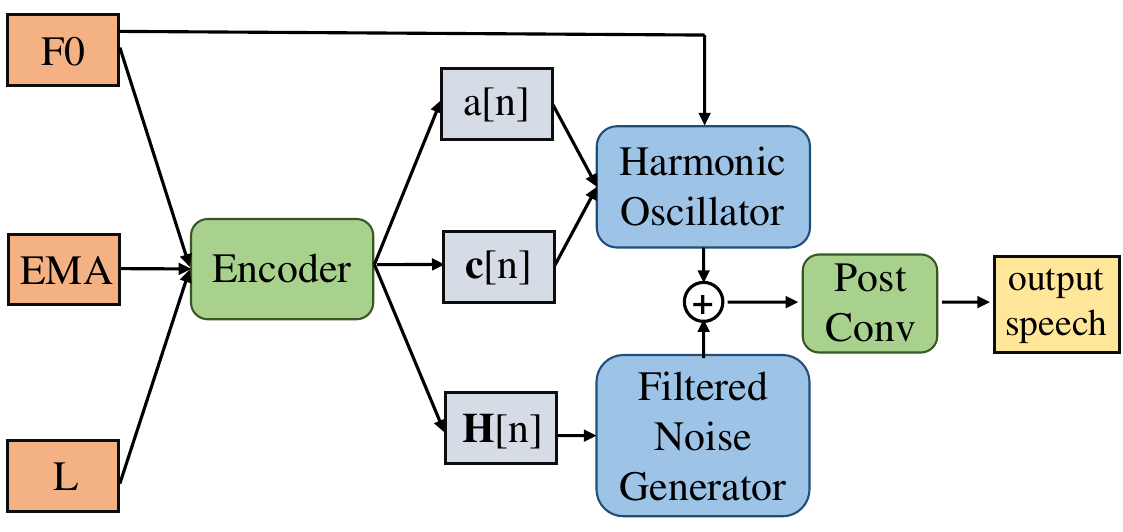}
    \caption{Overall model architecture. Only the green modules are trainable. The gray blocks are control signals. F0 is pitch, L is loudness, a[n] is the global amplitude, \textbf{c}[n] is the harmonic distribution, and \textbf{H}[n] is the filter frequency response.}
    \label{model}
\end{figure}
Following \cite{DDSP}, our proposed model mainly consists of two parts: an encoder and a DSP generator. The overall model architecture can be found in Figure \ref{model}. Note that F0 and loudness are pre-computed from the corresponding utterance. 

\subsection{Encoder}
\label{encoder}
\begin{figure}[t]
    \centering
    \includegraphics[scale=0.25]{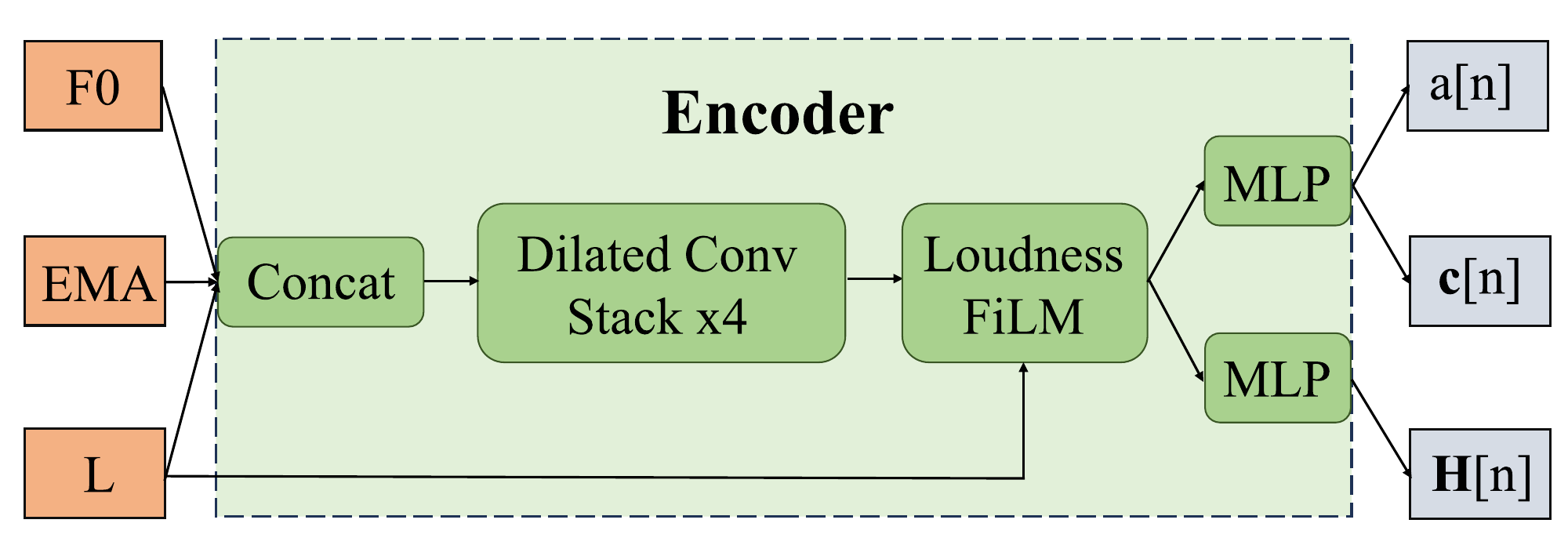}
    \caption{Encoder architecture. Loudness is fed into the loudness FiLM module as the condition.}
    \label{encoder architecture}
\end{figure}
The encoder architecture is shown in Figure \ref{encoder architecture}. Inspired by \cite{midi-ddsp}, we use a dilated convolution network as the encoder. The input to the encoder is F0, loudness, and EMA, all sampled at $f_{model}$ = 200Hz. The input features are first concatenated along the channel dimension, then processed by 4 dilated convolution stacks, while keeping the same time steps. In each stack there are 5 ResBlocks \cite{resnet} with dilations [1, 2, 4, 8, 16] respectively. The output is fed to a loudness conditioning FiLM \cite{film} layer, which takes in loudness as the condition and generates the affine transformation parameters to modulate the output features of dilated convolution stacks. FiLM helps to balance the amplitudes of harmonics and filtered noise, which will be mentioned in section \ref{dsp modules}.

The loudness FiLM output is processed by two multilayer perceptrons (MLPs). The first MLP produces $2(K+1)$-dimensional output: the first $K+1$ dimensions control sine waves, and the other $K+1$ control cosine waves. Each $K+1$ dimensional control signal comprises a global amplitude $a[n]$ and a $K$-dimensional time-varying harmonic distribution $\textbf{c}[n]$. Here, $K$ represents the total number of harmonics used. Harmonics exceeding the Nyquist frequency in $\textbf{c}[n]$ are set to -1e20 to avoid aliasing and then normalized via softmax. The other MLP output is the time-varying filter frequency response $\textbf{H}[n]$, an $M$-dimensional vector per time point $n$. To stabilize training, an exponential sigmoid nonlinearity, $\text{exp-sigmoid}(x) = 2.0 \cdot \text{sigmoid}(x)^{\log 10} + 10^{-7}$, is applied to $a[n]$ and $\textbf{H}[n]$, as per \cite{DDSP}.

\subsection{Digital Signal Processing (DSP) Generator}
\label{dsp modules}
For the DSP modules, we iterated on the DSP generators of \cite{DDSP}. The outputs of the encoder from section \ref{encoder} control two DSP modules: a harmonic oscillator and a filtered noise generator. The harmonic oscillator generates the voiced components of speech while the filtered noise generator synthesizes the unvoiced components. The outputs of these two modules are added to get the raw synthesized speech, which will be filtered by the post convolution (post conv) layer to generate the final synthesized speech.

\subsubsection{Harmonic Oscillator}
Unlike the harmonic oscillator in \cite{DDSP}, where only the sine harmonic waves are used, we propose to use both the corresponding sine harmonics and cosine harmonics to better approximate Fourier series for higher expressivity. The harmonic oscillator generates a sum of sine and cosine waves whose frequencies are multiples of F0. The $k$-th harmonic $x_k$ is controlled by global amplitudes $a[n], \Tilde{a}[n]$, harmonic weights $c_k[n], \Tilde{c}_k[n]$, and a frequency contour $f_k[n]$, as shown below in equation \ref{xk}.
\begin{equation}
    x_k[n] = a[n]c_k[n]\sin(\phi_k[n]) + \Tilde{a}[n]\Tilde{c} _k[n]\cos(\phi_k[n]) \label{xk}
\end{equation}
$\phi_k[n] = 2\pi\sum_{m=0}^{n}f_k[m]$ is the instantaneous phase and $f_k[n] = kF_0[n]$ is the integer multiple of F0. The harmonic distribution $\textbf{c}[n]$ (or $\Tilde{\textbf{c}}[n]$ for cosine waves) output from the encoder has $K$ values $(c_1[n], c_2[n], ..., c_K[n])^T$ for each time point $n$ and satisfy
\begin{equation}
    \sum_{k=0}^{K}c_k[n]=1 \quad \text{and} \quad c_k[n]\geq0
\end{equation}
Thus, the harmonic oscillator output can be calculated as
\begin{equation}
    x[n] = \sum_{k=1}^{K}(a[n]c_k[n]\sin(\phi_k[n]) + \Tilde{a}[n]\Tilde{c} _k[n]\cos(\phi_k[n]))
\end{equation}
Since $a[n], \Tilde{a}[n], F_0[n], \textbf{c}[n], \Tilde{\textbf{c}}[n]$ are all sampled at $f_{model}$ = 200Hz, we need to first upsample them back to the sampling frequency $f_s$ = 16kHz of the speech signals before calculating the above equations, i.e. upsample by a factor of $u$ = 80. Here $u$ is also the frame size. We upsample using the traditional signal-processing method by first inserting $u-1$ zeros between every two samples and then convolving with a Hann window of size $2u+1$. 

\subsubsection{Filtered Noise Generator}
This module generates noise signals filtered by learned linear time-varying finite impulse response (LTV-FIR) filters. To avoid complex numbers, we treat $\textbf{H}[n]$ as half of a zero-phase filter's transfer function, which is real and symmetric. We perform an inverse fast Fourier transform (FFT) to obtain zero-phase filter coefficients, shift them to form a causal, linear-phase filter, and apply a Hann window to balance time-frequency resolution, resulting in $\textbf{h}[n]$, which is then multiplied by an attenuation hyperparameter $\gamma$ to balance the filtered noise and harmonics. The filtered noise output is produced by convolving each $\textbf{h}[n]$ with a noise signal of length $u$ (a noise frame) and performing overlap-and-add with a hop size of $u$. Noise is generated from a uniform distribution between $[-1, 1]$, and all convolutions are computed via FFT. 

\subsection{Post Convolution Layer}
To further balance the noise and harmonics amplitudes, we introduce a post convolution (post conv) layer, which is a learnable 1D convolution layer without bias. Unlike the 1D convolution reverb module in \cite{DDSP}, which models reverberation or room acoustics, here the post conv layer acts as a filter to suppress the noise level or to compensate for the noise amplitudes depending on the previous amplitude balancing design choices. We explore this further in Section \ref{balance}.

\subsection{Loss Functions}

\subsubsection{Multi-Scale Spectral Loss}
\label{mss loss}
\begin{figure}[t]
    \centering
    \includegraphics[scale=0.37]{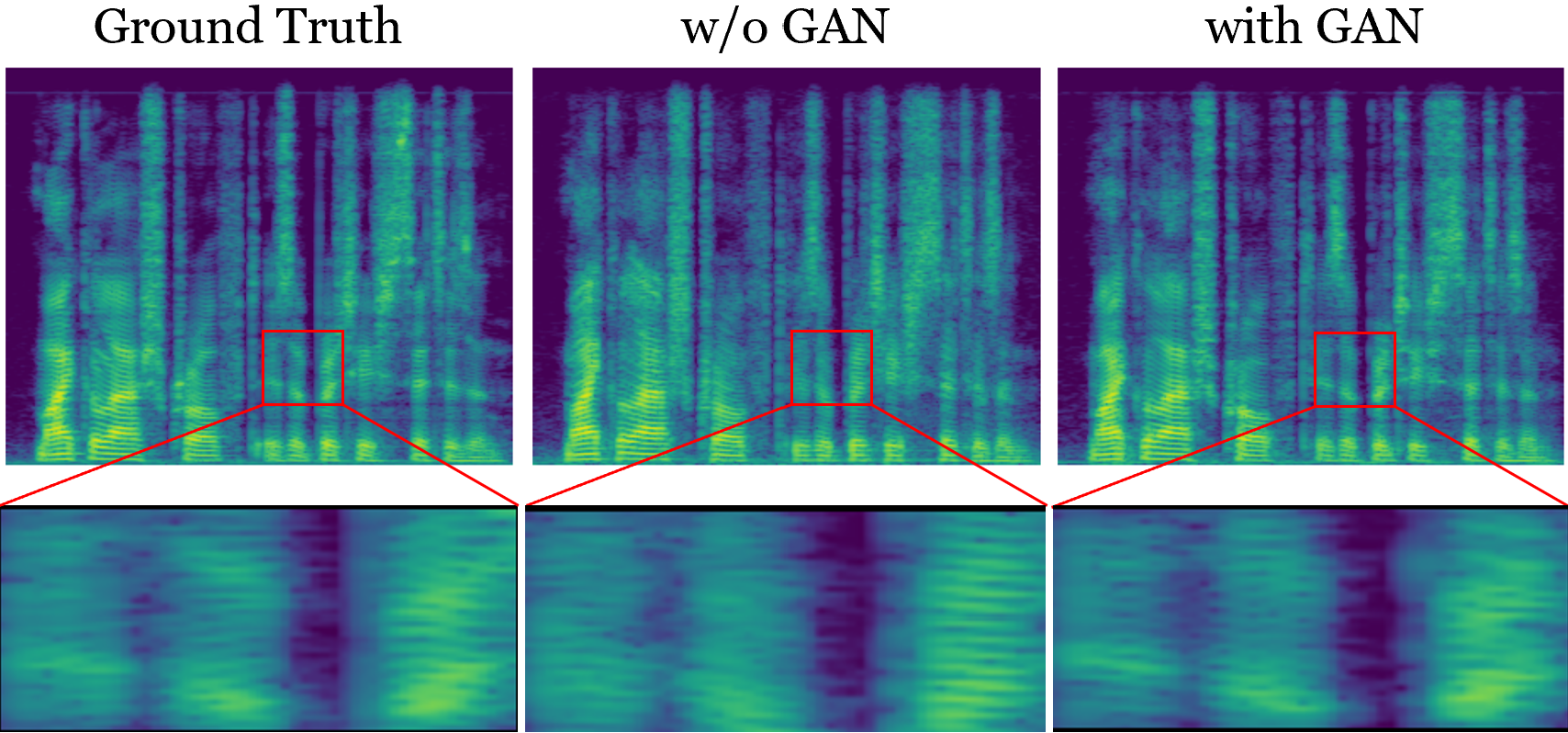}
    \caption{The spectrograms of the ground truth speech, synthesized speech without GAN, and synthesized speech with GAN. As shown in the boxed regions, without GAN the spectrogram energy bands are averaged out, while with GAN the finer structures are better preserved.}
    \label{muffled_sharp}
\end{figure} 

We use the multi-scale spectral loss as defined in \cite{DDSP}:
\begin{equation}
    \mathcal{L}_{MSS} = \sum_{i \in W} ||S_i - \hat{S}_i||_1 + \alpha ||\log S_i - \log \hat{S}_i||_1
\end{equation}
where $S$ and $\hat S$ are the magnitude spectrograms of the ground truth audio and the generated audio respectively. $\alpha$ is chosen to be $1$ in this paper. $W$ = [2048, 1024, 512, 256, 128, 64] is the set of FFT sizes, and the frame overlap is set to be $75\%$.

\subsubsection{Multi-Resolution Adversarial Loss}
\label{gan}


As mentioned in \cite{meta}, training only with multi-scale spectral loss for audio often results in over-smoothed spectrogram predictions. L1 / L2 losses aim to reduce large discrepancies and capture the low-frequency components of spectrograms, averaging out rapid changes in spectral details which results in muffled-sounding audio, as shown in Figure \ref{muffled_sharp}. 

To capture the finer details of spectrograms, following the work of \cite{MRD}, we utilize multi-resolution spectrogram discriminators. We treat each input spectrogram as a one-channel image, and perform 2D strided convolution for discrimination. Note that the input spectrograms are calculated from acoustics with different parameters, such as window size, hop size, and number of points for FFT, so that the discriminators have access to spectrograms of the same utterance with multiple resolutions.

For each sub-discriminator, the adversarial loss is calculated as Least Squares GAN (LSGAN) described in \cite{LSGAN}:

\begin{align}
    \min\limits_{D_i} \mathcal{L}_{LSGAN}(D_i;G) &= \frac{1}{2}\mathbb{E}_{x\sim p_{\text{data}}(x)}[(D_i(S(x)) - 1)^2] \nonumber \\
    &+ \frac{1}{2}\mathbb{E}_{z\sim p_z(z)}[(D_i(S(G(z))))^2]
\end{align}

\begin{equation}
    \min\limits_{G} \mathcal{L}_{LSGAN}(G;D_i) = \mathbb{E}_{z\sim p_z(z)}[(D_i(S(G(z))) - 1)^2]
\end{equation}
where $S$ is the magnitude STFT, $D_i$ is the i-th sub-discriminator, $G$ is the DDSP vocoder, $x$ is the ground truth audio, and $z$ is the input features. 

The loss functions for the generator and discriminator are:
\begin{equation}
    \mathcal{L}(G) = \mathcal{L}_{MSS} + \frac{\lambda}{R}\sum_{i=1}^{R}\mathcal{L}_{LSGAN}(G;D_i)
\end{equation}
\begin{equation}
    \mathcal{L}(D) = \frac{1}{R}\sum_{i=1}^{R}\mathcal{L}_{LSGAN}(D_i;G)
\end{equation}
where $R$ is the total number of sub-discriminators, which is also the total number of resolutions, and $\lambda$ controls the weight of the LSGAN loss.

\section{Results}

\begin{table*}[t]
\begin{center}
\caption{Model performance on MNGU0 and LJ Speech dataset. For UTMOS and MOS, the standard deviation is also reported.}
\label{performance}
\begin{tabular}{lccc|cc}
\toprule
Model Name & WER$\downarrow$ & PESQ$\uparrow$ & M-STFT$\downarrow$ & UTMOS$\uparrow$ & MOS$\uparrow$ \\
\midrule
Ground Truth (MNGU0) & 6.589 & - & - & 4.134 ± 0.170 & 3.910 ± 0.715 \\
HiFi-CAR (MNGU0) & 8.305 & 2.138 & 1.331 & 3.836 ± 0.202 & 3.575 ± 0.935 \\
DDSP (MNGU0) & \textbf{6.673} & \textbf{2.172} & \textbf{1.298} & \textbf{3.868 ± 0.182} & \textbf{3.735 ± 0.892} \\
\midrule
Ground Truth (LJ Speech) & 4.317 & - & - & 4.376 ± 0.123 & 4.165 ± 0.706 \\
HiFi-CAR (LJ Speech) & 4.557 & 1.962 & 1.296 & 3.795 ± 0.323 & 3.955 ± 0.838 \\
DDSP (LJ Speech) & \textbf{4.536} & \textbf{2.044} & \textbf{1.238} & \textbf{3.819 ± 0.332} & \textbf{4.025 ± 0.815} \\
\bottomrule
\end{tabular}
\end{center}
\end{table*}

\subsection{Datasets}
\label{dataset}
\subsubsection{MNGU0 EMA Dataset}
We experiment with the MNGU0 EMA dataset \cite{mngu0}, comprising 75 minutes of 16 kHz male speech with 200 Hz EMA recordings. The 12-dimensional EMA features capture the x and y coordinates of jaw, upper and lower lips, and tongue (tip, blade, and dorsum) movements. F0 is extracted from the speech using CREPE \cite{crepe} with a 5ms hop size, and loudness is computed as the maximum absolute amplitude of each 5ms speech frame \cite{loudness_1, loudness_2}. Consequently, EMA, F0, and loudness are all sampled at 200 Hz. During training, we randomly crop 1-second segments of aligned EMA, F0, and loudness for input, and their corresponding waveforms as targets. The dataset is split into 1129 training utterances (71.3 minutes) and 60 test samples (3.7 minutes), with 60 training utterances used for validation.

\subsubsection{LJ Speech Pseudo-Labelled Dataset}
To evaluate our model with a substantial amount of training data, we use the LJ Speech dataset \cite{ljspeech}, containing 24 hours of 22050 Hz female speech. As it lacks EMA data, we generate pseudo EMA labels using the acoustic-to-articulatory inversion (AAI) model from \cite{articulatory_encodec, inversion, ema_probing2}. EMA features are linearly interpolated from 50 Hz to 200 Hz, and waveforms are resampled to 16 kHz. Other features follow the MNGU0 settings. We use a 90\%/5\%/5\% train/validation/test split, corresponding to 21.5, 1.25, and 1.25 hours, respectively.

\subsection{Experimental Setup}
\label{setup}
For our DDSP model, we choose the kernel size of ResBlocks to be 3 with 2 convolution layers inside, the hidden dimension of the dilated convolution stacks to be 256, with \( K = 50 \) harmonics, \( M = 65 \) frequency bands, and attenuation \( \gamma = 0.01 \). The loudness FiLM module consists of three 1D convolution layers with kernel size 3, and the post convolution layer has a kernel size of 1025. This results in a total of 9.0M parameters. The multi-resolution discriminator uses \( R = 6 \) with FFT sizes [2048, 1024, 512, 256, 128, 64] and 75\% frame overlap. Weight normalization \cite{weightnorm} is applied to all sub-discriminators.

We use the Adam optimizer with \(\beta_1 = 0.9\), \(\beta_2 = 0.999\), and distinct learning rates: \(3 \times 10^{-4}\) for the generator and \(3 \times 10^{-6}\) for the discriminator. The batch size is 32, with \(\lambda = 5\). For MNGU0 dataset, there are 6400 training epochs. The learning rates are multiplied by $0.3$ at epoch milestones [2400, 4800]. For LJ Speech dataset, the total number of epochs is 1280, with epoch milestones = [480, 960]. The HiFi-CAR baseline (13.5M) \cite{Peter-ATS} is trained with its original configuration and adapted to our input features.

\subsection{Metrics}
We use both objective and subjective metrics to evaluate model performance. Objective metrics include: (1) word error rate (WER), which is calculated on the transcription of the synthesized test set speech using the SOTA speech recognition model Whisper-Large \cite{whisper}; A lower WER indicates higher intelligibility of the synthesized speech; (2) Multi-resolution STFT (M-STFT) \cite{PWG}\cite{auraloss}, which measures the difference between the spectrograms of the ground truth and the prediction across multiple resolutions; (3) perceptual evaluation of speech quality (PESQ) \cite{pesq}, a widely adopted automated method for assessing voice quality; and (4) UTMOS \cite{utmos}, a machine-evaluated mean opinion score (MOS). We use the conventional 5-scale MOS test as the subjective metric. Each model receives 200 unique ratings.

\subsection{Synthesis Quality}

The subjective and objective quality metrics for DDSP and HiFi-CAR are listed in Table \ref{performance}. For MNGU0, our DDSP model is consistently better than the baseline in every metric, with a boost in WER by 1.63$\%$ and a significant improvement in MOS (+0.16). This indicates that our DDSP model has a strong and appropriate inductive bias for the inner periodic structure of speech signals and is capable of generating high-fidelity speech. For LJ Speech, with substantially more training data, our model is still better in all metrics. This also indicates that our model is effectively compatible with the inverted EMA from the AAI model.

\subsection{Parameter Efficiency}

\begin{figure}[t]
    \centering
    \includegraphics[scale=0.47]{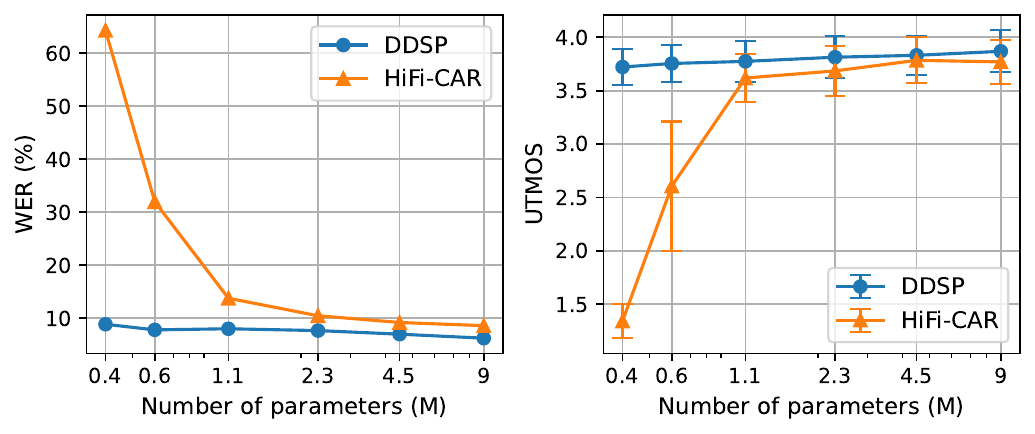}
    \caption{WER and UTMOS against model size.}
    \label{param-efficiency}
\end{figure} 

To evaluate parameter efficiency, we retrain the models using the same configurations as in section \ref{setup}, but with varying parameter counts (nparams): [9M, 4.5M, 2.3M, 1.1M, 0.6M, 0.4M]. For each nparams, we train the model with three random seeds (324, 928, 1024) and evaluate the combined synthesized test set speech. To maintain the receptive field size, we reduce nparams by decreasing the hidden dimension. The results are shown in Figure \ref{param-efficiency}. Our DDSP model shows no significant performance decline as model size decreases, outperforming HiFi-CAR at all nparams configurations. In contrast, HiFi-CAR's performance drops drastically below 1.1M nparams. Notably, our smallest model (0.4M) is comparable to HiFi-CAR (9M), highlighting our DDSP model's high parameter efficiency and potential for edge device applications.

\subsection{Inference Speed}

\begin{table}[t]
\begin{center}
\caption{Model sizes and CPU inference time for 1s of input.}
\label{speed}
\begin{tabular}{lrc}
\toprule
Model name & Params.$\downarrow$ & Inference Time [s]$\downarrow$ \\
\midrule
HiFi-CAR (13M) & 13.5M & 0.1805 ± 0.0480 \\
HiFi-CAR (9M) & 9.1M & 0.1445 ± 0.0117 \\
DDSP & \textbf{9.0M} & \textbf{0.0368 ± 0.0065} \\
\bottomrule
\end{tabular}
\end{center}
\end{table}

We test the inference speed of DDSP, HiFi-CAR (13M), and HiFi-CAR (9M) on an Apple M1 CPU by varying the input length \(N\) from 0.5s to 10s, with 0.5s intervals. For each \(N\), we average the inference time over 50 utterances of the same length \(N\), normalizing by \(N\). Table \ref{speed} reports the model sizes and the mean and standard deviation of the average inference time for 1s of input. Our model is 1.5x smaller and 4.9x faster than HiFi-CAR (13M). Notably, HiFi-CAR (9M) is still 3.9x slower than DDSP despite having the same model size. Furthermore, as shown in Figure \ref{param-efficiency}, HiFi-CAR (9M) consistently underperforms compared to DDSP in both WER and UTMOS. This demonstrates that our DDSP model is fast and lightweight without sacrificing synthesis quality.

\subsection{Ablation Study}

\begin{table*}[t]
\begin{center}
\caption{Ablation study. }
\label{ablation}
\begin{tabular}{lccc|cc}
\toprule
Model Name & WER$\downarrow$ & PESQ$\uparrow$ & M-STFT$\downarrow$ & UTMOS$\uparrow$ & MOS$\uparrow$ \\
\midrule
DDSP (MNGU0) & 6.673 & 2.172 & 1.298 & 3.868 ± 0.182 & 3.735 ± 0.892 \\
\midrule
w/o GAN loss & 6.064 & 2.279 & 1.264 & 2.129 ± 0.435 & 3.100 ± 1.020 \\
w/o cosine & 10.379 & 1.823 & 1.410 & 3.090 ± 0.214 & 3.295 ± 1.038 \\
w/o post conv & 10.335 & 1.940 & 1.364 & 3.426 ± 0.179 & 3.460 ± 0.974 \\
w/o FiLM & 7.375 & 2.100 & 1.315 & 3.829 ± 0.195 & 3.485 ± 0.905 \\
\bottomrule
\end{tabular}
\end{center}
\end{table*}

We perform an ablation study on the GAN loss, additional cosine harmonics, post conv layer, and loudness FiLM using the MNGU0 dataset, with all models trained under the same configuration as the original model. The results, summarized in Table \ref{ablation}, show that removing any module decreases performance, except for the GAN loss. Without the GAN loss, similarity metrics like PESQ and M-STFT improve, as the model is trained solely on reconstruction loss ($\mathcal{L}_{MSS}$ in Section \ref{mss loss}), leading to predictions more similar to the ground truth on average but perceptually over-smoothed, as mentioned in Section \ref{gan} and supported by significant drops in UTMOS (-1.739) and MOS (-0.64). The absence of additional cosine harmonics causes substantial performance drops across all metrics, underscoring their importance in speech modeling. The post conv layer is essential for balancing noise and harmonics amplitudes. Omitting the loudness FiLM module results in a small yet noticeable performance decline.

\section{Discussion}
\label{discussion}
\subsection{Speech Decomposition}

\begin{figure}[t]
    \centering
    \includegraphics[scale=0.17]{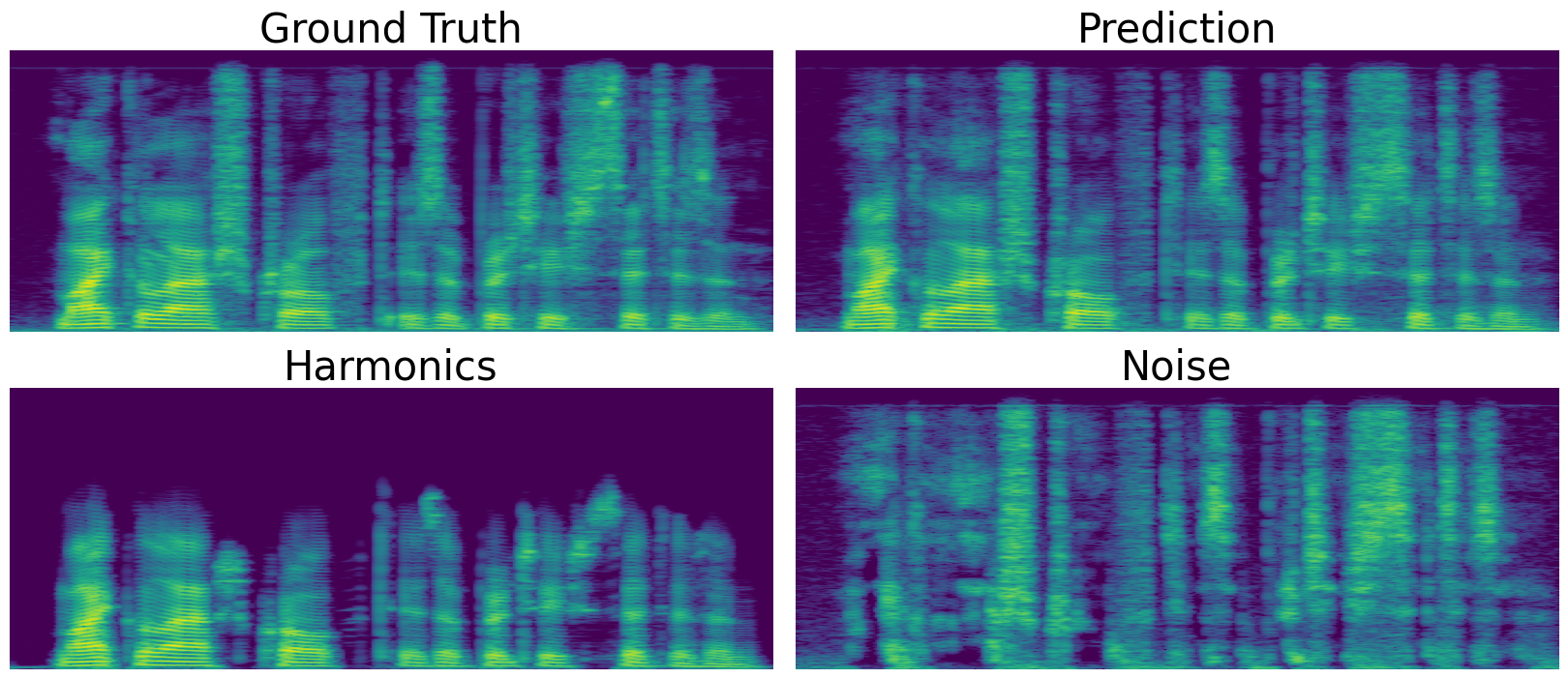}
    \caption{Decomposed spectrograms of the utterance ``Michael Ashcroft is a British citizen."}
    \label{decomposition}
\end{figure} 

Since the synthesized speech is the sum of harmonics and filtered noise, we can decompose the output and visualize each component via spectrograms (Figure \ref{decomposition}). The harmonics spectrogram shows distinct frequency bands and higher energy, reflecting the quasi-periodic nature of voiced sounds generated by the harmonic oscillator. In contrast, the noise spectrogram displays higher frequency components with a dispersed energy distribution along the frequency axis, modeling the unvoiced, noise-like sounds such as fricatives and consonants.

\subsection{Noise / Harmonics Balance}
\label{balance}
\begin{figure}[t]
    \centering
    \includegraphics[scale=0.35]{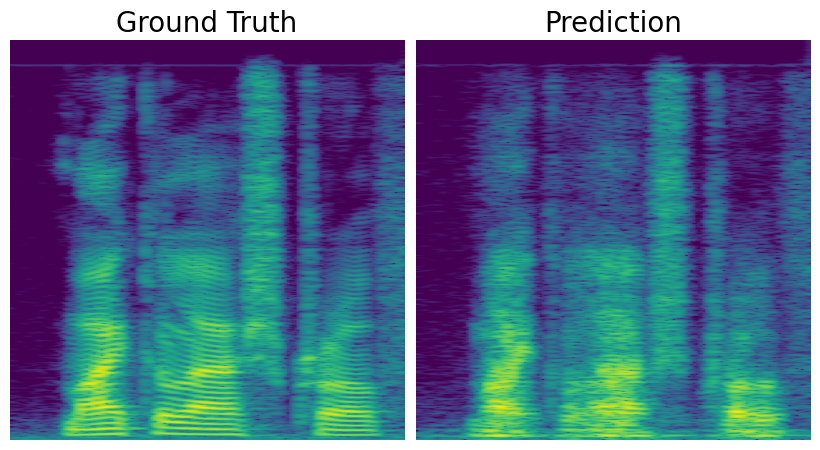}
    \caption{The spectrograms of the ground truth and the amplitude-imbalanced audio. The predicted spectrogram has lost all harmonic structures.}
    \label{imbalance}
\end{figure} 

\begin{figure}[t]
    \centering
    \includegraphics[scale=0.35]{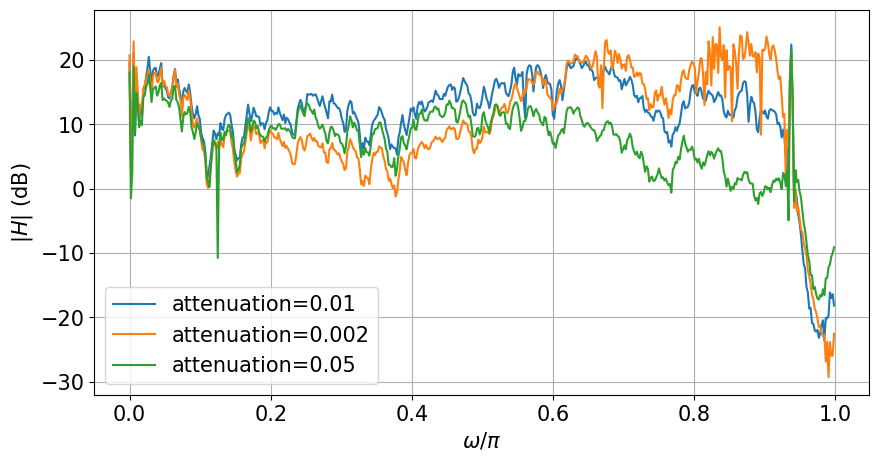}
    \caption{The frequency responses of the learned post conv filters using different attenuation hyperparameters. }
    \label{attenuation}
\end{figure} 
One challenge in achieving a high-quality vocoder using our DDSP model is balancing the amplitudes of harmonics and noise. We employ three methods to address this issue: the attenuation hyperparameter $\gamma$, the post conv layer, and the loudness FiLM module. Among these methods, the attenuation and the post conv layer are particularly crucial. If there is no attenuation at all, i.e. $\gamma=1$, the model will only learn the filtered noise as shown in Figure \ref{imbalance}. Although on average the energy distribution seems correct, the predicted spectrogram has lost all finer harmonic structures, while for the ground truth, there are clear and detailed harmonic stripes. 

We have also analyzed the frequency responses of the learned post conv filters when trained with different levels of attenuation, as shown in Figure \ref{attenuation}. The attenuation parameter $\gamma$ influences the noise energy: higher $\gamma$ results in greater noise amplitude. Given that harmonic energy is concentrated in the lower frequencies while noise has high energy in the higher frequency range, the post conv filter should suppress high-frequency components to balance the noise and harmonics amplitudes when $\gamma$ is large. This is evidenced in Figure \ref{attenuation}, where the gain $|H|$ decreases in the high-frequency range ($\omega > 0.4\pi$) as $\gamma$ increases. This demonstrates that the attenuation and post conv filter together effectively balance the noise and harmonics amplitudes.

\section{Conclusion}
In this paper, we present a DDSP articulatory vocoder based on harmonic-plus-noise model. With the strong inductive bias of DDSP, we show that our model is parameter-efficient, fast, and capable of synthesizing high-quality speech from EMA, F0 and loudness. For future work, we plan to explore the multi-speaker capabilities of our DDSP vocoder. 

\section{Acknowledgements}
This research is supported by the following grants to PI Anumanchipalli: NSF award 2106928, Google Research Scholar Award, Rose Hills Foundation and UC Noyce Foundation.
\newpage

\bibliographystyle{IEEEbib}
\bibliography{strings,refs}

\end{document}